\documentclass[twocolumn]{aastex631}

\submitjournal{APJ}

\begin{document}

\title{SDSS-IV MaNGA: Physical Origins of Double-Peaked Narrow Emission-Line Spaxels in Barred  Galaxies}

\author[0000-0001-9410-9485]{Jiajie Qiu}
\affiliation{Shanghai Astronomical Observatory, Chinese Academy of Sciences, 80 Nandan Rd., Shanghai, 200030, China}
\affiliation{School of Astronomy and Space Science, University of Chinese Academy of Sciences, 1 East Yanqi Lake Rd., Beijing 100049, P.R. China}
\affiliation{Key Lab for Astrophysics, Shanghai, 200034, China}

\author[0000-0002-3073-5871]{Shiyin Shen}
\affiliation{Shanghai Astronomical Observatory, Chinese Academy of Sciences, 80 Nandan Rd., Shanghai, 200030, China}
\affiliation{Key Lab for Astrophysics, Shanghai, 200034, China}

\author[0000-0002-8733-1587]{Ruixiang Chang}
\affiliation{Shanghai Astronomical Observatory, Chinese Academy of Sciences, 80 Nandan Rd., Shanghai, 200030, China}
\affiliation{Key Lab for Astrophysics, Shanghai, 200034, China}

\author{Qianwen Zhao}
\affiliation{Shanghai Astronomical Observatory, Chinese Academy of Sciences, 80 Nandan Rd., Shanghai, 200030, China}
\affiliation{School of Astronomy and Space Science, University of Chinese Academy of Sciences, 1 East Yanqi Lake Rd., Beijing 100049, P.R. China}
\affiliation{Key Lab for Astrophysics, Shanghai, 200034, China}
\author{Qi Zeng}

\affiliation{Shanghai Astronomical Observatory, Chinese Academy of Sciences, 80 Nandan Rd., Shanghai, 200030, China}
\affiliation{School of Astronomy and Space Science, University of Chinese Academy of Sciences, 1 East Yanqi Lake Rd., Beijing 100049, P.R. China}
\affiliation{Key Lab for Astrophysics, Shanghai, 200034, China}

\correspondingauthor{Shiyin Shen}
\email{ssy@shao.ac.cn}

\begin{abstract}
The physical origins of double-peaked narrow emission-line spaxels (DPSs) in barred galaxies are explored through the analysis of a sample of 72 barred double-peaked emission-line galaxies (DPGs) extracted from the MaNGA dataset. In this study, we examine two potential scenarios: the gas inflow along the bar and the formation of a bar-induced gaseous nuclear ring. By applying a classical galactic dynamics model, we calculate the radii and rotational velocities of the nuclear rings for all barred DPGs, and compare them with the observed properties of their DPSs. Our analysis reveals a significant correlation between the predicted radii of the nuclear rings and the maximum centric distances of the DPSs, as well as a marginal correlation between the predicted rotational velocities of the nuclear rings and the observed maximum velocity differences of the DPSs. These findings provide strong evidence to support the hypothesis that the DPSs of a barred DPG in MaNGA primarily originate from the convolution of the PSF effect with its bar-induced fast-rotating gaseous nuclear ring.
\end{abstract}


\section{Introduction\label{sec:Introduction}}

The optical emission lines of galaxies are of immense importance in resolving the physical properties of their ionized gas \citep{Springel2005, Benson2010, Rubinur2016}. Typically, each galaxy exhibits only one set of narrow emission lines, whose velocity is consistent with the average motion of the galaxy itself. However, observations have identified galaxies exhibiting multiple sets of emission lines \citep{Heckman1981, Heckman1984}, suggesting the presence of additional components with motions that differ significantly from their average motions. The galaxies exhibiting two sets of narrow emission lines, designated as the double-peaked narrow emission-line galaxies \citep[DPGs,][]{Xu2009}, have been explored by many spectroscopic surveys. It has been determined that these DPGs do not constitute a single category of galaxies with analogous properties, but originated from different physical mechanisms \citep{Ge2012}. For instance, by kinematically classifying the centric emission lines, \citet{Nevin2016} proposed the mechanisms that could produce double-peaked features include kpc-scale dual AGNs, biconical outflows of ionized gas driven by AGNs, bar-induced inflows, and rotating disks, etc.

As initiated by \citet{Wang2018}, the physical origins of DPGs are deduced using the integral-field spectroscopic (IFS) data, in which a ring-like structure is identified for the galaxy (MaNGA-ID 1-556501) in the SDSS MaNGA survey \citep{Abazajian2009, Bundy2014}. Subsequently, \citet{Ciraulo2021} identified a target (MaNGA-ID 1-114955) that exhibited double-peaked features, which were revealed to be a merging system consisting of two disks overlapping. And \citet{Yazeedi2021} presented a study of the galaxy (MaNGA-ID 1-166919), demonstrating its association with the bi-conical outflows of its AGN.

A recent census of the double-peaked features of the H$ \alpha $-[N \uppercase\expandafter{\romannumeral2}]$ \lambda\lambda $ 6549, 6586 \AA\ emission lines of all spaxels in the final data release of the MaNGA survey has been published by \citet{Qiu2024}, where a sample of 5,420 double-peaked narrow emission-line spaxels (DPSs) hosted by 304 DPGs has been identified. By further associating with the physical properties of their host galaxies, these DPSs are statistically divided into three different categories, corresponding to three different physical mechanisms: bars, tidal interactions, and AGN activities. The majority of the DPSs, located in the inner regions of galaxies ($r/R_{\rm{e}} \sim 0.17$, where $R_{\rm{e}}$ is the effective radius of its host galaxy) and exhibit a minor velocity difference ($\Delta v \sim 225$ km/s, between two emission-line peaks), are statistically associated with the bar structures of galaxies.

It has been posited that the double-peaked features of barred galaxies may be attributable to the actions of multiple physical mechanisms. Firstly, bars serve as the channels of gas inflows, allowing rapid movement from the outskirts to galaxy centers \citep{Geron2023, Feng2024, Kim2024}, which may generate the secondary set of emission lines when the bars are oriented along the line-of-sight \citep{Maschmann2023}. Secondly, the gaseous nuclear rings, which have been suggested as the coherent structures of bars that accumulate gas around the inner Lindblad Resonance \citep[ILR,][]{Lindblad1963, Aswathy2020, Sormani2024}. \citet{Sparke2006} demonstrated that the unresolved gaseous ring in NGC 7331 produces double-horned emission lines. In a similar manner, the emission line profiles for the IFS observation with limited spatial resolution would exhibit a double-peaked shape and vary according to the spatial location of the spaxels \citep{Schmidt2019, Maschmann2023}. 

Despite the proposal of several mechanisms, the interaction of bars with the kinematics of galaxies remains poorly understood, particularly with regard to its effects on gas inflows and the formation of nuclear rings \citep{Liu2025}. Consequently, elucidating the physical origins of the double-peaked features in barred galaxies is instrumental in fostering a more profound comprehension of the associated dynamical processes. The following is structured as follows. The construction of the sample of barred DPGs is outlined in Section \ref{sec:Data}. In Section \ref{sec:Methods and Results}, we assess the two hypotheses of the physical origins of the DPSs in the barred galaxies in detail. In Section \ref{sec:Discussion}, we offer concise discussions on the barred galaxies without DPSs in MaNGA. Finally, we draw conclusions in Section \ref{sec:Conclusions}.

\section{Data\label{sec:Data}}

We use the barred DPG sample from \citet{Qiu2024}, which contains 304 DPGs from the final data release of MaNGA survey. Since the DPSs in a DPG may result from multiple physical processes (e.g., bars, AGNs or tidal interactions), we take the DPGs with only bar features, i.e. these galaxies without any AGN or tidal interaction. This selection yields a sample of 72 barred DPGs, containing 1,068 DPSs.

We take the stellar mass ($M_*$) of these DPGs and their central SNRs of H$\alpha$ emission lines ($SNR_{\rm{H\alpha}}$) from the MaNGA pipe3D Catalog \citep{Sanchez2016}. For the DPSs, we extract the $r/R_{\rm{e}}$ and $\Delta v$ from \citet{Qiu2024}.

\section{Methods and Results\label{sec:Methods and Results}}

\subsection{Origin of DPSs: Gas Inflows along Bars}

If the DPSs in barred galaxies were originated from the ionized gas inflows along the bar directions, we would expect that their galactic disks to exhibit high inclinations and their bars to be preferentially aligned perpendicular to the major axes of the galaxies \citep{Maschmann2023}. 

\subsubsection{Inclinations of Disks of Barred DPGs\label{sec:Inclination of Disks}}

\begin{figure*}
\centering
\includegraphics[width=1\textwidth]{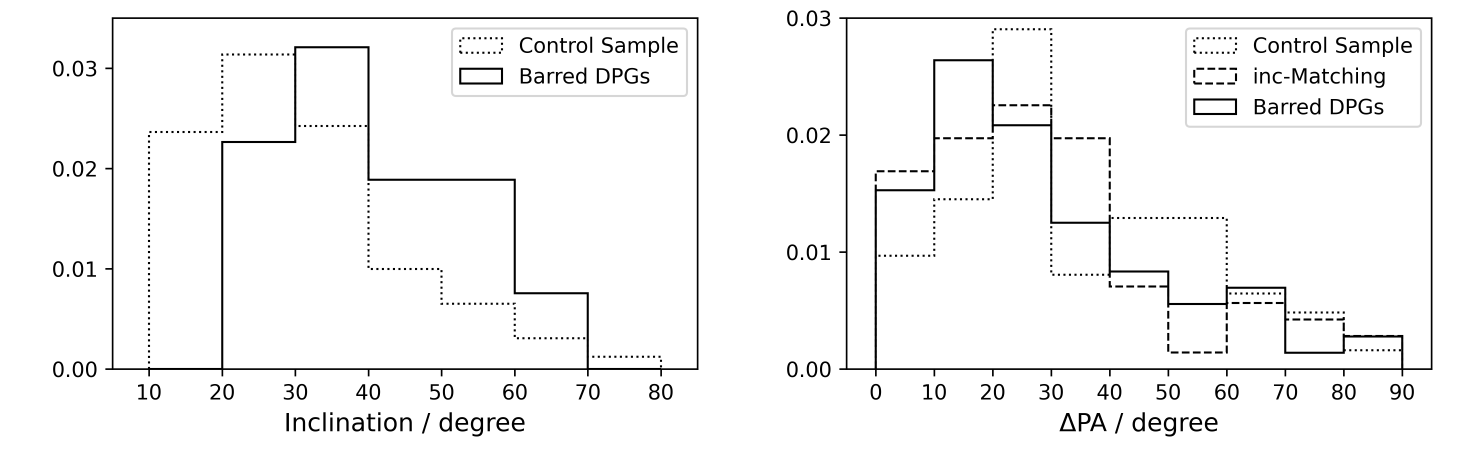}
\caption{\label{fig:gas_inflows}
The distributions of the disk inclinations $inc$ (left) and bar orientations $\Delta PA$ (right) for 72 barred DPGs and the control samples of barred galaxies without DPSs. 
The p-value of $\Delta PA$ distributions in the Kolmogorov-Smirnov test for the barred DPGs and $inc$-control sample is 0.88.}
\end{figure*}

We take the isophotal $b/a$ in $r$ bands of the MaNGA galaxies from NSA Catalog \citep{Maller2009}, and their inclinations ($inc$) of the disks are calculated through 
\begin{equation}
\sin(inc) = \frac{1 - (b/a)}{1 - (b/a)_{\rm{min}}}\label{inclination}\,
\end{equation}
where $(b/a)_{\rm{min}} = 0.15$ corresponds to the minimum axis ratio of the edge-on galaxies ($inc = 90^{\circ}$).

The left panel of Fig. \ref{fig:gas_inflows} shows the $inc$ distribution of our barred DPGs (solid histogram). It is biased toward low inclinations, possibly due to a selection effect that the bars in face-on galaxies are more easily identified by visual inspection. To quantify this effect, we select a control sample of barred galaxies lacking DPSs in MaNGA, but with their $M_*$ and central $SNR_{\rm{H\alpha}}$ individually matched in a one-to-one manner to exclude the possibility that non-detections of DPSs due to low ionized gas densities. The distribution of $inc$ for this control sample is shown in the left panel as the dotted histogram, suggesting that the control sample of barred galaxies is also biased to face-on galaxies. Comparing barred galaxies with and without DPSs, it appears that barred DPGs are generally more inclined. This observation supports the hypothesis that DPS features are linked to gas dynamics within the disk planes, which makes such movements easier to detect at higher inclinations. However, this plane motion does not directly indicate gas movement along the bar, prompting further exploration into the alignment of bars within the galactic plane.

\subsubsection{Alignment of Bars on Disk Planes\label{sec:alignment}}

To characterize the alignment of the bars on the galactic plane, we locate the full bar shapes in galaxy images using bar masks from the Galaxy Zoo: 3D project \citep{Masters2021}. The bar regions of all barred galaxies were examined on the synthesized $r$ images (525 $\times$ 525 pixels, with a scale of 0.099 arcsec per pixel) by at least 15 volunteers. Each pixel of the galaxy images is assigned a value between 0 and 15, indicating the number of volunteers who identified the pixel as part of a bar. Based on these bar masks, we visually select all the pixels with mask values greater than 7.5 within a minimized rectangle, and measure the position angle ($PA_{\rm{bar}}$) as well as the bar length ($R_{\rm{bar}}$) of the rectangle's long side. Then, we use the Python PAFIT package \citep{Krajnovic2006} to calculate the position angle ($PA$) of the kinematic major axis of the host galaxy from its stellar velocity field by the MaNGA Data Analysis Pipeline \citep[DAP,][]{Westfall2019}. Finally, we calculate the alignment of a bar on the galactic plane by $\Delta PA = |PA_{\rm{bar}}-PA|$. 

The right panel of Fig. \ref{fig:gas_inflows} illustrates the $\Delta PA$ distributions for the DPGs and their control sample, which match $M_*$ and $SNR_{H\alpha}$, as constructed in Section \ref{sec:Inclination of Disks}, shown by the solid and dotted histograms, respectively. Compared to the control sample, the $\Delta PA$ values of barred DPGs are skewed towards higher values. This trend is likely linked to the disk inclination bias of the barred DPGs, shown in the left panel. For highly inclined disks, if the bar is perpendicular to the major axis (high $\Delta PA$), the projected bar length may be too short for visual detection. To investigate this bias, we create a new control sample of normal barred galaxies without DPSs, using $M_*$, $SNR_{\rm{H\alpha}}$, and $inc$ as control parameters. This new sample is called the $inc$-control sample. The dashed histogram in the right panel represents the $\Delta PA$ distribution for this $inc$-control sample. Once $inc$ is controlled, the $\Delta PA$ distribution of barred DPGs aligns closely with that of normal barred galaxies, as confirmed by the Kolmogorov-Smirnov test with a p-value of 0.85. In other words, we find no evidence that one of the velocity components in the DPSs of the barred DPGs represents the ionized gas inflow along the bar.

\subsection{Origin of DPSs: Gaseous Nuclear Rings\label{sec:Nuclear Rings}}

In this section, we test the hypothesis that the DPS features in barred galaxies are linked to the nuclear rings \citep{Schmidt2019}.

\subsubsection{Nuclear Rings of Barred DPGs\label{sec:Nuclear Rings of Barred DPGs}}

\begin{figure*}
\centering
\includegraphics[width=1\textwidth]{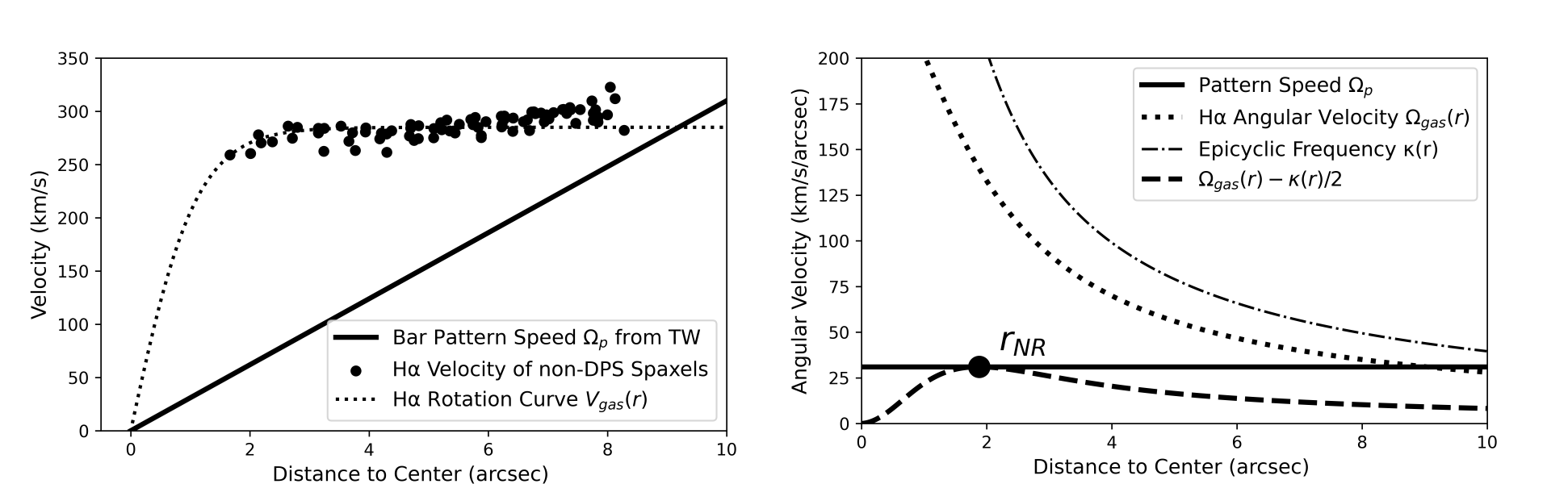}
\caption{\label{fig:rc}
Left: The pattern speed of the bar and the rotation curves of the example barred DPG (MaNGA-ID 1-593159). The solid line represents the pattern speed ($\Omega_{\rm{p}}$) of the bar obtained via TW algorithm. The black filled circles indicate the positions and H$\alpha$ velocities of the non-DPS spaxels on the major axis from MaNGA DAP, and the dotted line represents the gas rotation curve $V_{\rm{gas}}(r)$.
Right: The pattern speed of the bar ($\Omega_{\rm{p}}$) and the radius of the gaseous nuclear ring ($r_{\rm{NR}}$) for the same DPG. The dotted line is the gas angular velocity $\Omega_{\rm{gas}}(r)$ and the dash-dotted line is the epicyclic frequency $\kappa(r)$. The radius of the gaseous nuclear ring is indicated using a filled circle where $\Omega_{\rm{p}}$ (solid line) equal to $\Omega_{\rm{gas}}(r) - \kappa(r)/2$ (dashed line).
}
\end{figure*}

In the context of the standard galactic dynamics model, the bar motion generates an inner Lindblad resonance, where it then forms a nuclear ring structure \citep{Aswathy2020}. The radius of this nuclear ring (denoted as $r_{\rm{NR}}$), can be inferred from the bar's pattern speed ($\Omega_{\rm{p}}$) and the angular velocity of the gas $\Omega_{\rm{gas}}(r) = V_{\rm{gas}}(r)/r$, where $V_{\rm{gas}}(r)$ is the gas rotation curve. At the radius $r = r_{\rm{NR}}$, the following equation is satisfied:
\begin{equation}
\Omega_{\rm{p}} = \Omega_{\rm{gas}}(r) - \kappa(r)/2, \label{ilr}
\end{equation}
where $\kappa$ denotes the epicyclic frequency of the gas in the radial direction \citep{Schmidt2019} and is characterized by the equation \ref{kappa}:
\begin{equation}
\kappa^2(r)= 4\Omega^2_{\rm{gas}}(r)\left[1+\frac{1}{2}\left(\frac{\mathit{r}}{\Omega_{\rm{gas}}(\mathit{r})}\frac{\rm{d}\Omega_{\rm{gas}}(r)}{\rm{d}\mathit{r}}\right)\right] \,.\label{kappa}
\end{equation}

We use the Tremaine-Weinberg (TW) algorithm \citep{Tremaine1984} to calculate $\Omega_{\rm{p}}$ of the bars, which is a model-independent approach and works for the galaxies with $20^{\circ}<inc < 70^{\circ}$. A detailed example of calculation of $\Omega_{\rm{p}}$ of the barred galaxies in MaNGA can be found in Section 2 in \citet{Geron2023}. In this study, we follow the same algorithm \footnote{More details see https://zenodo.org/records/7567945}.

To obtain the $V_{\rm{gas}}(r)$ of each barred DPG, we first select spaxels (no DPS) with SNR \footnote{SNR is calculated by the method in \citet{Qiu2024}} $ > 5$ in the H$\alpha$ velocity fields (from DAP) on the major axes of the DPGs, and then use the Python package scipy.optimize.curve\_fit \citep{Virtanen2020} \footnote{https://docs.scipy.org/doc/scipy/reference/generated /scipy.optimize.curve\_fit.html} to obtain the $V_{\rm{gas}}(r)$ with Equation \ref{gas rotation curve}:
\begin{equation}
V_{\rm{gas}}(r) = V_{\rm{gas,0}} \times \tanh(\frac{r}{r_{\rm{gas,0}}}), \label{gas rotation curve}
\end{equation}
where $r_{\rm{gas,0}}$ and $V_{\rm{gas,0}}$ represent the transition radius and asymptotic velocity of the gas rotation curve, respectively. Besides the simple tanh function, other analytical models have been suggested for fitting the rotation curve of disk galaxies, such as the classical C97 model \citep{Courteau1997}. In fact, the specific results of this study are largely unaffected by the specific form of the function used to fit the rotational curve. A detailed comparison between the C97 and tanh models is shown in Appendix \ref{sec:C97}.

We use the standard $\chi^2$ fitting technic to fit the model parameters $V_{\rm{gas,0}}$ and $r_{\rm{gas,0}}$ for each rotation curve. The left panel of Fig. \ref{fig:rc} shows the $V_{\rm{gas}}(r)$ measurement for the example DPG (MaNGA-ID 1-593159), where the H$\alpha$ velocities of non-DPSs obtained directly from DAP data are shown as the solid dots and the fitted $V_{\rm{gas}}(r)$ curve is shown as the dotted line.

Given that DPSs appear mainly at the centers of barred galaxies, non-DPS spaxels may not be able to cover the rising part of the rotation curve. This would lead to biases in the fitting of the model parameter $r_{\rm{gas,0}}$. For our barred DPGs, we find that 13 galaxies fall into this category, referred to as the "flat sample" below. For them, we adopt the maximum possible value of $r_{\rm{gas,0}}$. In Appendix \ref{sec:flat}, we present the details of the fitting process of three specific examples: a normal sample (MaNGA-ID 1-195478), a "flat sample" (MaNGA-ID 1-10248), and the example DPG.

After determining the values of $\Omega_{\rm{p}}$ and $V_{\rm{gas}}(r)$, we calculate the radius of the inner Lindblad resonances ($r_{\rm{NR}}$) by combining the Equations \ref{ilr} and \ref{kappa}.
The right panel of Fig. \ref{fig:rc} illustrates the calculation of $r_{\rm{NR}}$ for the example DPG, depicting $\Omega_{\rm{gas}}(r)$ and $\kappa(r)$ as the dotted and dash-dotted lines, respectively. The dashed curve represents $\Omega_{\rm{gas}}(r)-\kappa(r)/2$, and the horizontal line denotes $\Omega_{\rm{p}}$. The point where these lines intersect corresponds to the radius $r_{\rm{NR}}$ (the example DPG has $r_{\rm{NR}} = 1^{\prime\prime}.9$). 
For different galaxies, the number of intersections between $\Omega_{\rm{gas}}(r)-\kappa(r)/2$ curve and $\Omega_{\rm{p}}$ line can be one, two, or none. In cases where there are two intersections, the outer intersection is selected as $r_{\rm{NR}}$. For 19 galaxies lacking $r_{\rm{NR}}$, we exclude them from further analysis. We exclude the 19 galaxies lacking $r_{\rm{NR}}$ from further analysis. Of the 72 barred DPGs, 53 $r_{\rm{NR}}$ values were found to be situated between $0^{\prime\prime}.8$ and $2^{\prime\prime}.2$. 

Upon determining $r_{\rm{NR}}$, we use the fitted gas rotation curve $V_{\rm{gas}}(r_{\rm{NR}})$ to calculate the rotation velocities ($v_{\rm{NR}}$) of the gas on nuclear rings (the example DPG has $v_{\rm{NR}} = 268$ km/s).
It is important to note that determining the $v_{\rm{NR}}$ for DPGs directly from the H$\alpha$ velocity field maps supplied by DAP is not feasible. This is because the $r_{\rm{NR}}$ values are similar to the PSF scale (FWHM $\sim 2^{\prime\prime}.5$) of the MaNGA survey. As a result, the H$\alpha$ velocities observed within the rings result from convolving the true velocity map with the PSF \citep{Maschmann2023}. This very effect could cause the DPG phenomena, which we will explore in detail below.

\begin{figure*}
\centering
\includegraphics[width=1\textwidth]{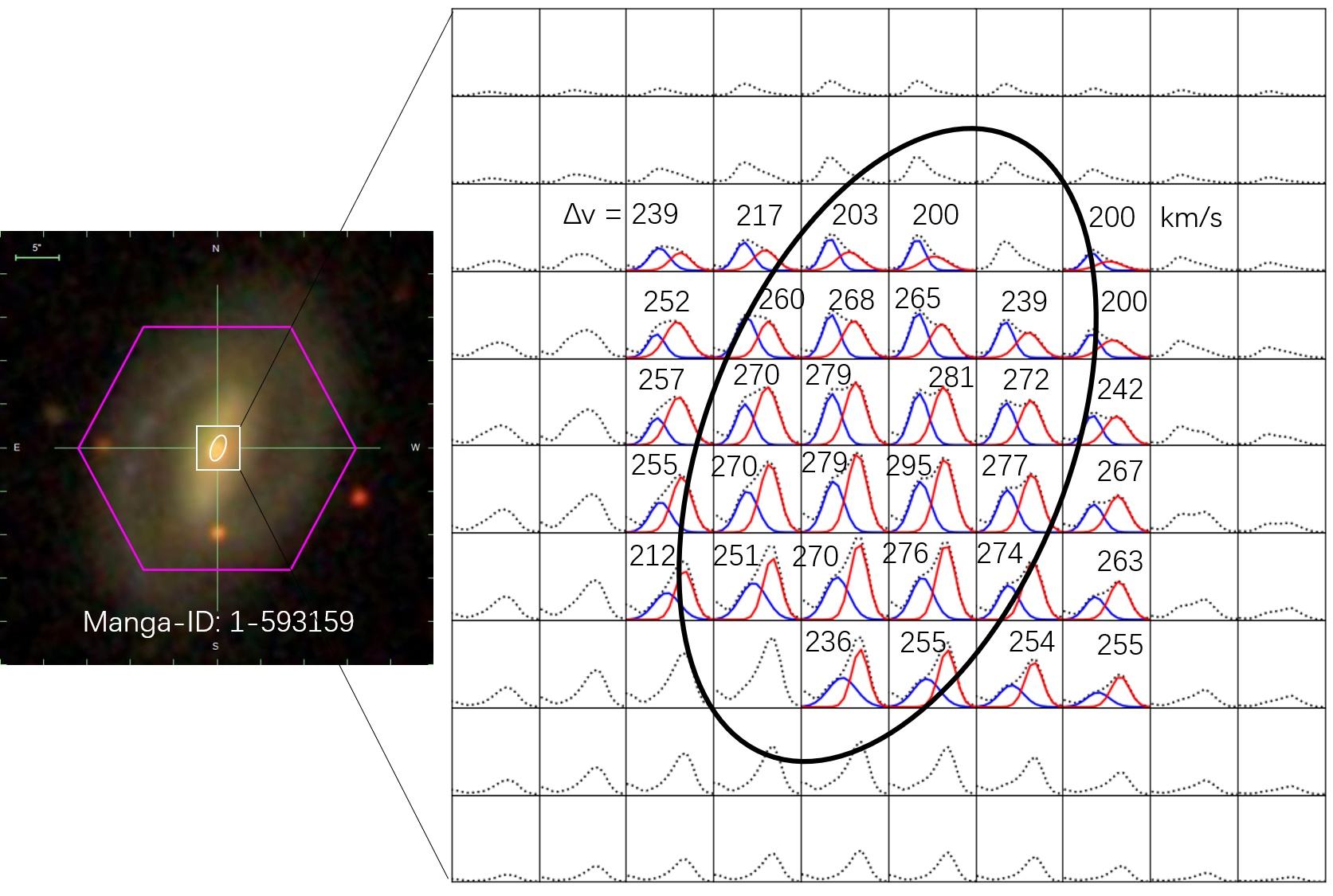}
\caption{\label{fig:example}
The SDSS $g,r,i$ band image for the example barred DPG is shown in the left panel. The white box is its central $10 \times 10$ spaxels ($5^{\prime\prime} \times 5^{\prime\prime}$), and the white ellipse is the inclined gaseous nuclear ring with $inc = 38^{\circ}$. The right panel shows the H$ \alpha $ line profiles of its central $10 \times 10$ spaxels. Spaxels fitted by the double Gaussian profiles are shown with the blue- and red-shifted H$ \alpha $ lines additionally, plotted by the blue and red lines respectively. The value labeled in each DPS is its $\Delta v$ value in km/s. The black ellipse is the gaseous nuclear ring ($r_{\rm{NR}}=1^{\prime\prime}.9$ and $\Delta v_{\rm{NR}}=330$ km/s) obtained from dynamical modelling in section \ref{sec:Comparsion}.
}
\end{figure*}

\subsubsection{Comparsion Observed DPS Features with Modelled Nuclear Rings \label{sec:Comparsion}}

\begin{figure*}
\centering
\includegraphics[width=1\textwidth]{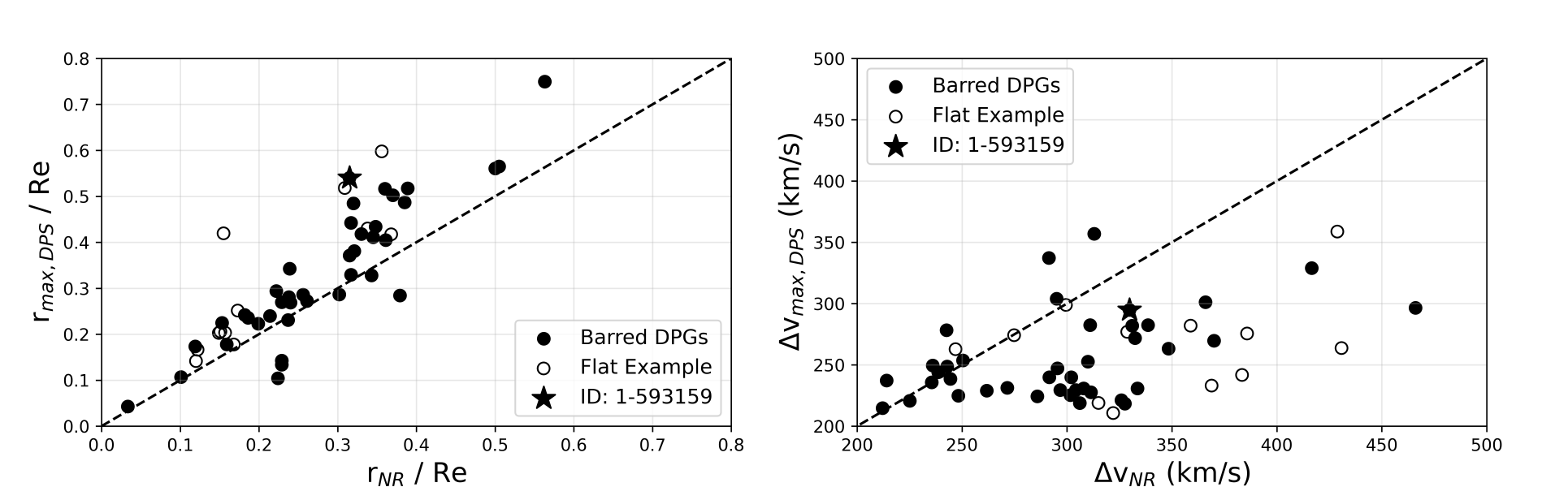}
\caption{\label{fig:nr}
The correlation between the $r_{\rm{NR}}$ and $r_{\rm{max,DPS}}$ of the barred DPGs is shown in the left panel and that between the $\Delta v_{\rm{NR}}$ and $v_{\rm{max,DPS}}$ is shown in the right panel. The  "flat samples" are labeled by circles and the example DPG is denoted by a pentagram. The dashed line in each panels is the one-to-one diagonal.
}
\end{figure*}

Before matching the observed DPS characteristics with the modeled nuclear rings in barred galaxies, it is necessary to determine how the DPS features in the MaNGA data are associated with the modeled nuclear rings.

We show the DPSs of the example barred DPG in Fig. \ref{fig:example}, where the nuclear ring is shown by the ellipse ($r_{\rm{NR}} = 1^{\prime\prime}.9$, $inc = 38^{\circ}$). It is evident that the DPSs span an area comparable to that of the nuclear ring, supporting the hypothesis that they arise from the PSF convolution of the nuclear ring's velocity structure. Consequently, we anticipate that the maximum central radial distance of the DPSs, $r_{\rm{max,DPS}}$, in barred DPGs will closely correlate with $r_{\rm{NR}}$. 

For each DPS, we also label the velocity difference $\Delta v$ of the two peaks in the corner of the panels in Fig. \ref{fig:example}. The $\Delta v$ values are in the range from $200-300$ km/s, with the maximum value appearing on the innermost spaxel. The velocity profile for the double-peaked emission lines of a given spaxel is influenced by the ratio of redshifted and blueshifted velocity components from the rotating gas entering the spaxel. Therefore, accurately modeling $\Delta v$ for a specific spaxel requires more than just understanding the rotation velocity of the nuclear ring, $\Delta v_{\rm{NR}}$; comprehensive modeling of the ionized gas distribution and PSF effect is necessary. However, this level of detailed modeling is beyond the scope of this study.
In a simplified scenario where all the ionized gas is on the nuclear ring, the greatest velocity difference between the two velocity components appears at the center, as described in \citet{Sparke2006}
\begin{equation}
\Delta v_{\rm{NR}} = 2v_{\rm{NR}} \times \sin(inc), \label{conversion}
\end{equation}
In a more realistic situation where ionized gas is also found within the radius of the nuclear ring, the $\Delta v_{\rm{NR}}$ serves as the maximum possible $\Delta v$ value measurable from DPSs, aligning with our example DPG that shows $\Delta v_{\rm{max}}$ (295 km/s) slightly less than $\Delta v_{\rm{NR}}$ (330 km/s).

Keeping in mind the arguments above about the correlation between nuclear rings and DPS properties for the example DPG, we calculate $r_{\rm{max,DPS}}$ and $\Delta v_{\rm{max,DPS}}$ for all 53 barred DPGs for which their $r_{\rm{NR}}$ have been successfully measured. We then compare them to $r_{\rm{NR}}$ and $\Delta v_{\rm{NR}}$ respectively, as illustrated in Fig. \ref{fig:nr}.
The left panel presents a comparison of $r_{\rm{max,DPS}}$ and $r_{\rm{NR}}$, with both radii being normalized by their host galaxy's effective radius. We observe a tight correlation between $r_{\rm{max,DPS}}$ and $r_{\rm{NR}}$ with a Pearson correlation coefficient of 0.90 (0.82 if “flat samples” are included). Moreover, as indicated by the diagonal line in the plot, this relationship nearly follows a 1:1 ratio. This finding strongly suggests that DPS features in barred DPGs are physically associated with nuclear rings. 
The right panel of Fig. \ref{fig:nr} illustrates the comparison between $\Delta v_{\rm{max,DPS}}$ and $\Delta v_{\rm{NR}}$. As predicted by the nuclear ring model, $\Delta v_{\rm{NR}}$ is typically greater than all values of $\Delta v_{\rm{max,DPS}}$. Furthermore, $\Delta v_{\rm{NR}}$ shows a slight positive correlation with $\Delta v_{\rm{max,DPS}}$ (with a Pearson correlation coefficient of 0.46, or 0.40 if “flat samples” are included), which further supports the hypothesis that the nuclear rings are the central origin of the DPS phenomena observed in barred galaxies.

\section{Discussion\label{sec:Discussion}}

\subsection{Other Properties of Bars}

\begin{figure*}
\centering
\includegraphics[width=1\textwidth]{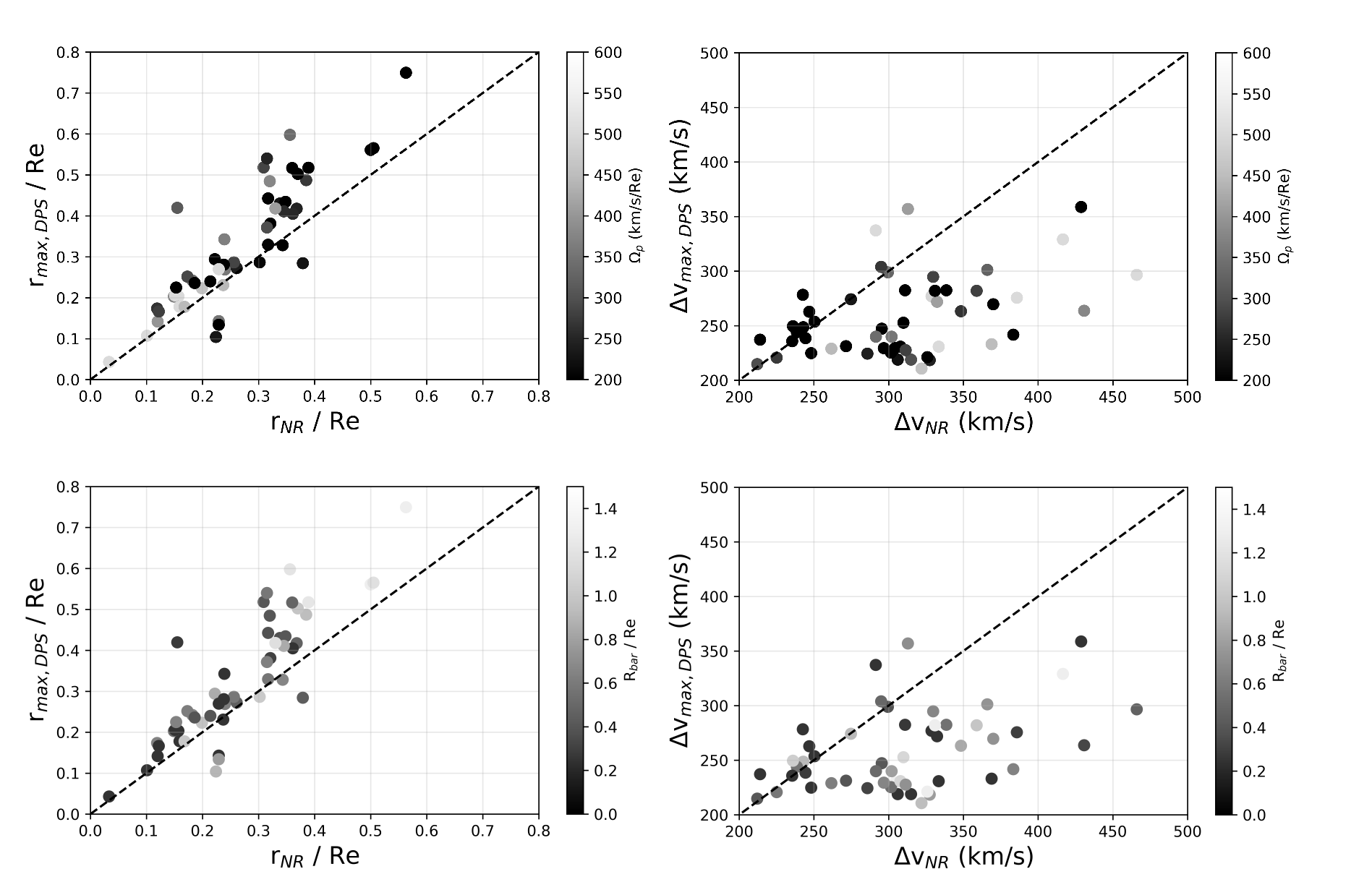}
\caption{\label{fig:vp}
The same relations as those in Fig. \ref{fig:nr}, but the color of the points now represents the values of the bar $\Omega_{\rm{p}}$ (top panels) and $R_{\rm{bar}}$ (bottom panels) repectively.
}
\end{figure*}

In this study, we have obtained significant correlations between the observed DPS features and the modeled nuclear ring properties for all barred DPGs available in MaNGA. Whether these correlations show  dependences on other physical properties of bars? We check the dependences on two available physical properties of the bars, the pattern speed ($\Omega_{\rm{p}}$ in section \ref{sec:Nuclear Rings of Barred DPGs}) and length ($R_{\rm{bar}}$ in section \ref{sec:alignment}) of the bar. The results are shown in Fig. \ref{fig:vp}.
From these panels, we do not find any significant dependences on these two physical properties of bars for either the correlation between $r_{\rm{NR}}$ and  $r_{\rm{max,DPS}}$ or $\Delta V_{\rm{NR}}$ and $\Delta v_{\rm{max,DPS}}$. Nevertheless, it is also possible that our sample of barred DPGs is too small to reveal some second-order correlation effects.

\begin{figure*}
\centering
\includegraphics[width=1\textwidth]{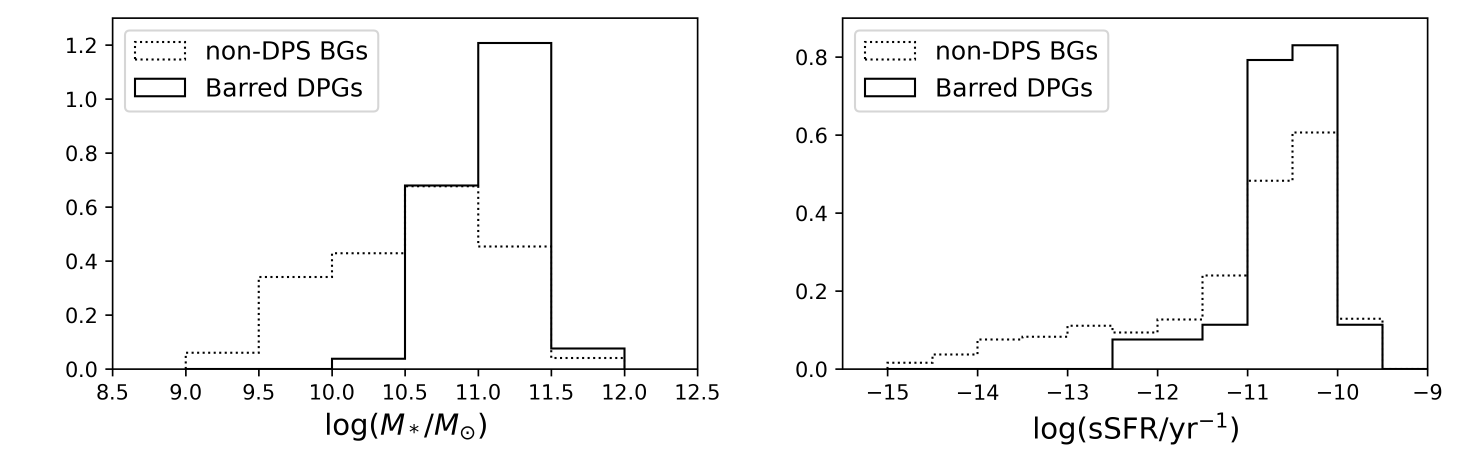}
\caption{\label{fig:mssfr} The distributions of $\rm{M_*}$ (left) and $\rm{sSFR}$ (right) of the barred DPGs (solid histograms) and all non-DPS barred galaxies (dotted) in MaNGA.
}
\end{figure*}

\subsection{Barred Galaxies without DPSs in MaNGA}

In section \ref{sec:Comparsion}, we illustrated a causal connection from bars to nuclear rings and subsequently to observed DPS features. However, it's important to note that the proportion of barred galaxies exhibiting DPS features in MaNGA is relatively small.

According to the Galaxy Zoo 2 Catalog \citep{Willett2013}, the MaNGA survey includes 1,151 barred galaxies, 123 of which have been identified as DPGs \citep{Qiu2024}. As we have shown in Section \ref{sec:Data}, a subset of 72 DPGs is associated with bar features only. This is the sample being probed in this study. In Section \ref{sec:Nuclear Rings}, we successfully calculated  the radii of the nuclear rings for 53 out of the 72 bar-only DPGs.  Within the MaNGA sample, DPGs tend to have higher $M_*$ and $sSFR$ \citep{Qiu2024}, mainly due to the DPS identification process, which requires a minimum signal-to-noise ratio for emission lines (closely linked to $sSFR$) and a minimum velocity difference ($\Delta v_{min} > $ 200\ km/s, relating to $M_*$). As a result, the 72 barred DPGs exhibit $M_*$ within the interval of $\log (\rm{M_*} / \rm{M_\odot}) > 10.5$, with $sSFR$ of $\log (\rm{sSFR} / \rm{yr^{-1}}) > -12.5$. The MaNGA dataset includes 519 barred galaxies that fall within the same parameter ranges but lack any DPS features as identified by the algorithm in \citet{Qiu2024}.

First, we compare the $M_*$ and $SNR_{H\alpha}$ distributions of the 72 barred DPGs and the 519 non-DPS barred galaxies in Fig. \ref{fig:mssfr}. As shown in the figure, compared to non-DPS barred galaxies, even after considering the minimum $M_*$ cut, the barred DPGs are significantly more biased toward massive galaxies. This bias in $M_*$ is likely due to the faster increase of the rotation curves at the centers of the massive galaxies, making nuclear rings easier to be detected as double-peaked shapes \citep{Patel2024}. On the other hand, after considering the minimum $sSFR$ cut, the barred DPGs and non-DPS barred galaxies show similar $sSFR$ distributions, with the barred DPGs showing slightly higher $sSFR$. The slight bias in $sSFR$ is likely due to the higher fraction of cold gas in galaxies with higher $sSFR$, where the emission line features of the nuclear ring  then are more prominent.

Moreover, the barred galaxies without the DPS feature are also related to the disk inclination. As shown in the left panel of Fig. \ref{fig:gas_inflows}, we have shown that the barred DPGs are significantly biased toward more inclined than the controlled non-DPS galaxies. For these barred galaxies with low inclinations, the $\Delta v_{\rm{NR}}$ would be too small to detect as double-peaked features in emission lines (see Equation \ref{conversion}).

Finally, the barred galaxies with or without DPS feature is likely correlated with the bar strength. Strongly barred galaxies typically exhibit strong non-circular motions \citep{Kim2024}, which manifest as increased rotation velocities of the gas in nuclear rings \citep{Liu2025}. Checking the bar strength (strong or weak bar) in the Galaxy Zoo 2 catalog \citep{Willett2013}, reveals that the strong bar fraction of barred DPGs is as high as 63\%, while the strong bar fraction of control non-DPS galaxies is only 40\%. However, the connection between the bar strength and the rotational velocity of nuclear ring is complicated by the fact that multiple dynamical mechanisms influence the gas close to the center of the galaxy simultaneously \citep{Athanassoula2009}.

In summary, the bar structure can trigger the formation of a nuclear ring. Under certain observational conditions, the emission lines from the ionized gas in the nuclear ring can exhibit double-peaked features. These conditions may be too strict for the MaNGA dataset, resulting in most barred galaxies in MaNGA not showing clearly detected DPS features.

\section{Conclusions\label{sec:Conclusions}}

This study investigates the physical origins of DPS features in barred galaxies using a sample of 72 barred DPGs with DPSs, extracted from a larger set of 304 DPGs as documented in \citet{Qiu2024}. We examine two possible scenarios: the gas inflows along the bars and the formation of bar-induced gaseous nuclear rings. Regarding the former, we evaluate the alignment of bars within the disk plane and discover that DPG bars are not significantly oriented perpendicular to the disk's major axis, which otherwise could have served as statistical evidence supporting the gas inflow hypothesis. For the latter, we use a classical galactic dynamics model to calculate the radii and rotational velocity of the bar-induced nuclear rings for each barred DPG, and compare them with the observed properties of the DPSs in these DPGs. We find a significant correlation between the maximum centric distances of the DPSs and the radii of the gaseous nuclear rings, and  also a marginal correlation between the maximum velocity differences of the double peaks in the DPSs and the theoretical velocity differences produced by gaseous nuclear rings. These novel findings provide strong evidence that the double-peaked features of these barred DPGs mainly originate from the convolution of a PSF effect and a fast-rotating nuclear ring in MaNGA's integrated-field spectroscopy. Our work opens a new avenue for studying the nuclear rings of barred galaxies.

Furthermore, utilizing IFS instruments with improved spatial resolution, like CSST-IFS, to study the central regions of barred DPGs can provide a deeper understanding of the physical properties of nuclear rings (Feng et al. in submit).

\section*{ACKNOWLEGEMENTS}

SS thanks research grants from the China Manned Space Project with NO. CMS-CSST-2025-A07, the National Key R\&D Program of China (No. 2022YFF0503402), Shanghai Academic/Technology Research Leader (22XD1404200) and National Natural Science Foundation of China (No. 12073059).

Funding for the Sloan Digital Sky Survey IV has been provided by the Alfred P. Sloan Foundation, the U.S. Department of Energy Office of Science, and the Participating Institutions. SDSS-IV acknowledges support and resources from the Center for High Performance Computing at the University of Utah. The SDSS website is www.sdss.org.

\bibliography{reference}{}

\begin{thebibliography}{}
\expandafter\ifx\csname natexlab\endcsname\relax\def\natexlab#1{#1}\fi
\providecommand{\url}[1]{\href{#1}{#1}}
\providecommand{\dodoi}[1]{doi:~\href{http://doi.org/#1}{\nolinkurl{#1}}}
\providecommand{\doeprint}[1]{\href{http://ascl.net/#1}{\nolinkurl{http://ascl.net/#1}}}
\providecommand{\doarXiv}[1]{\href{https://arxiv.org/abs/#1}{\nolinkurl{https://arxiv.org/abs/#1}}}

\bibitem[{{Abdurro'uf} {et~al.}(2009){Abdurro'uf}, {Accetta}, {Aerts}, {Silva Aguirre}, {Ahumada}, {Ajgaonkar}, {Filiz Ak}, {Alam}, {Allende Prieto}, {Almeida}, {Anders}, {Anderson}, {Andrews}, {Anguiano}, {Aquino-Ort{\'\i}z}, {Arag{\'o}n-Salamanca}, {Argudo-Fern{\'a}ndez}, {Ata}, {Aubert}, {Avila-Reese}, {Badenes}, {Barb{\'a}}, {Barger}, {Barrera-Ballesteros}, {Beaton}, {Beers}, {Belfiore}, {Bender}, {Bernardi}, {Bershady}, {Beutler}, {Bidin}, {Bird}, {Bizyaev}, {Blanc}, {Blanton}, {Boardman}, {Bolton}, {Boquien}, {Borissova}, {Bovy}, {Brandt}, {Brown}, {Brownstein}, {Brusa}, {Buchner}, {Bundy}, {Burchett}, {Bureau}, {Burgasser}, {Cabang}, {Campbell}, {Cappellari}, {Carlberg}, {Wanderley}, {Carrera}, {Cash}, {Chen}, {Chen}, {Cherinka}, {Chiappini}, {Choi}, {Chojnowski}, {Chung}, {Clerc}, {Cohen}, {Comerford}, {Comparat}, {da Costa}, {Covey}, {Crane}, {Cruz-Gonzalez}, {Culhane}, {Cunha}, {Dai}, {Damke}, {Darling}, {Davidson}, {Davies}, {Dawson}, {De Lee}, {Diamond-Stanic}, {Cano-D{\'\i}az}, {S{\'a}nchez},
  {Donor}, {Duckworth}, {Dwelly}, {Eisenstein}, {Elsworth}, {Emsellem}, {Eracleous}, {Escoffier}, {Fan}, {Farr}, {Feng}, {Fern{\'a}ndez-Trincado}, {Feuillet}, {Filipp}, {Fillingham}, {Frinchaboy}, {Fromenteau}, {Galbany}, {Garc{\'\i}a}, {Garc{\'\i}a-Hern{\'a}ndez}, {Ge}, {Geisler}, {Gelfand}, {G{\'e}ron}, {Gibson}, {Goddy}, {Godoy-Rivera}, {Grabowski}, {Green}, {Greener}, {Grier}, {Griffith}, {Guo}, {Guy}, {Hadjara}, {Harding}, {Hasselquist}, {Hayes}, {Hearty}, {Hern{\'a}ndez}, {Hill}, {Hogg}, {Holtzman}, {Horta}, {Hsieh}, {Hsu}, {Hsu}, {Huber}, {Huertas-Company}, {Hutchinson}, {Hwang}, {Ibarra-Medel}, {Chitham}, {Ilha}, {Imig}, {Jaekle}, {Jayasinghe}, {Ji}, {Johnson}, {Jones}, {J{\"o}nsson}, {Katkov}, {Khalatyan}, {Kinemuchi}, {Kisku}, {Knapen}, {Kneib}, {Kollmeier}, {Kong}, {Kounkel}, {Kreckel}, {Krishnarao}, {Lacerna}, {Lane}, {Langgin}, {Lavender}, {Law}, {Lazarz}, {Leung}, {Leung}, {Lewis}, {Li}, {Li}, {Lian}, {Liang}, {Lin}, {Lin}, {Lin}, {Lintott}, {Long}, {Longa-Pe{\~n}a}, {L{\'o}pez-Cob{\'a}}, {Lu},
  {Lundgren}, {Luo}, {Mackereth}, {de la Macorra}, {Mahadevan}, {Majewski}, {Manchado}, {Mandeville}, {Maraston}, {Margalef-Bentabol}, {Masseron}, {Masters}, {Mathur}, {McDermid}, {Mckay}, {Merloni}, {Merrifield}, {Meszaros}, {Miglio}, {Di Mille}, {Minniti}, {Minsley}, {Monachesi}, {Moon}, {Mosser}, {Mulchaey}, {Muna}, {Mu{\~n}oz}, {Myers}, {Myers}, {Nadathur}, {Nair}, {Nandra}, {Neumann}, {Newman}, {Nidever}, {Nikakhtar}, {Nitschelm}, {O'Connell}, {Garma-Oehmichen}, {Luan Souza de Oliveira}, {Olney}, {Oravetz}, {Ortigoza-Urdaneta}, {Osorio}, {Otter}, {Pace}, {Padilla}, {Pan}, {Pan}, {Parikh}, {Parker}, {Peirani}, {Pe{\~n}a Ram{\'\i}rez}, {Penny}, {Percival}, {Perez-Fournon}, {Pinsonneault}, {Poidevin}, {Poovelil}, {Price-Whelan}, {B{\'a}rbara de Andrade Queiroz}, {Raddick}, {Ray}, {Rembold}, {Riddle}, {Riffel}, {Riffel}, {Rix}, {Robin}, {Rodr{\'\i}guez-Puebla}, {Roman-Lopes}, {Rom{\'a}n-Z{\'u}{\~n}iga}, {Rose}, {Ross}, {Rossi}, {Rubin}, {Salvato}, {S{\'a}nchez}, {S{\'a}nchez-Gallego}, {Sanderson}, {Santana
  Rojas}, {Sarceno}, {Sarmiento}, {Sayres}, {Sazonova}, {Schaefer}, {Schiavon}, {Schlegel}, {Schneider}, {Schultheis}, {Schwope}, {Serenelli}, {Serna}, {Shao}, {Shapiro}, {Sharma}, {Shen}, {Shetrone}, {Shu}, {Simon}, {Skrutskie}, {Smethurst}, {Smith}, {Sobeck}, {Spoo}, {Sprague}, {Stark}, {Stassun}, {Steinmetz}, {Stello}, {Stone-Martinez}, {Storchi-Bergmann}, {Stringfellow}, {Stutz}, {Su}, {Taghizadeh-Popp}, {Talbot}, {Tayar}, {Telles}, {Teske}, {Thakar}, {Theissen}, {Tkachenko}, {Thomas}, {Tojeiro}, {Hernandez Toledo}, {Troup}, {Trump}, {Trussler}, {Turner}, {Tuttle}, {Unda-Sanzana}, {V{\'a}zquez-Mata}, {Valentini}, {Valenzuela}, {Vargas-Gonz{\'a}lez}, {Vargas-Maga{\~n}a}, {Alfaro}, {Villanova}, {Vincenzo}, {Wake}, {Warfield}, {Washington}, {Weaver}, {Weijmans}, {Weinberg}, {Weiss}, {Westfall}, {Wild}, {Wilde}, {Wilson}, {Wilson}, {Wilson}, {Wolf}, {Wood-Vasey}, {Yan}, {Zamora}, {Zasowski}, {Zhang}, {Zhao}, {Zheng}, {Zheng}, \& {Zhu}}]{Abazajian2009}
{Abdurro'uf}, {Accetta}, K., {Aerts}, C., {et~al.} 2009, The Astrophysical Journal Supplement Series, 182, 543, \dodoi{10.1088/0067-0049/182/2/543}

\bibitem[{Aswathy \& Ravikumar(2020)}]{Aswathy2020}
Aswathy, S., \& Ravikumar, C.~D. 2020, Research in Astronomy and Astrophysics, 20, 015, \dodoi{10.1088/1674-4527/20/2/15}

\bibitem[{Athanassoula {et~al.}(2009)Athanassoula, Romero-Gómez, \& Masdemont}]{Athanassoula2009}
Athanassoula, E., Romero-Gómez, M., \& Masdemont, J.~J. 2009, Monthly Notices of the Royal Astronomical Society, 394, 67, \dodoi{10.1111/j.1365-2966.2008.14273.x}

\bibitem[{Benson(2010)}]{Benson2010}
Benson, A.~J. 2010, Physics Reports, 495, 33, \dodoi{https://doi.org/10.1016/j.physrep.2010.06.001}

\bibitem[{Bundy {et~al.}(2014)Bundy, Bershady, Law, Yan, Drory, MacDonald, Wake, Cherinka, S{\'{a}}nchez-Gallego, Weijmans, Thomas, Tremonti, Masters, Coccato, Diamond-Stanic, Arag{\'{o}}n-Salamanca, Avila-Reese, Badenes, Falc{\'{o}}n-Barroso, Belfiore, Bizyaev, Blanc, Bland-Hawthorn, Blanton, Brownstein, Byler, Cappellari, Conroy, Dutton, Emsellem, Etherington, Frinchaboy, Fu, Gunn, Harding, Johnston, Kauffmann, Kinemuchi, Klaene, Knapen, Leauthaud, Li, Lin, Maiolino, Malanushenko, Malanushenko, Mao, Maraston, McDermid, Merrifield, Nichol, Oravetz, Pan, Parejko, Sanchez, Schlegel, Simmons, Steele, Steinmetz, Thanjavur, Thompson, Tinker, van~den Bosch, Westfall, Wilkinson, Wright, Xiao, \& Zhang}]{Bundy2014}
Bundy, K., Bershady, M.~A., Law, D.~R., {et~al.} 2014, The Astrophysical Journal, 798, 7, \dodoi{10.1088/0004-637x/798/1/7}

\bibitem[{Ciraulo {et~al.}(2021)Ciraulo, Melchior, Maschmann, Katkov, Halle, Combes, Gelfand, \& {Al Yazeedi}}]{Ciraulo2021}
Ciraulo, B., Melchior, A., Maschmann, D., {et~al.} 2021, Astronomy and Astrophysics, 653, \dodoi{10.1051/0004-6361/202141319}

\bibitem[{Courteau(1997)}]{Courteau1997}
Courteau, S. 1997, Astron. J., 114, 2402, \dodoi{10.1086/118656}

\bibitem[{Feng {et~al.}(2024)Feng, Li, Shen, Gerhard, Saglia, Blaña, Li, \& Jing}]{Feng2024}
Feng, Z.-X., Li, Z., Shen, J., {et~al.} 2024, The Astrophysical Journal, 963, 22, \dodoi{10.3847/1538-4357/ad13ee}

\bibitem[{Ge {et~al.}(2012)Ge, Hu, Wang, Bai, \& Zhang}]{Ge2012}
Ge, J.-Q., Hu, C., Wang, J.-M., Bai, J.-M., \& Zhang, S. 2012, The Astrophysical Journal Supplement Series, 201, 31, \dodoi{10.1088/0067-0049/201/2/31}

\bibitem[{Géron {et~al.}(2023)Géron, Smethurst, Lintott, Kruk, Masters, Simmons, Mantha, Walmsley, Garma-Oehmichen, Drory, \& Lane}]{Geron2023}
Géron, T., Smethurst, R.~J., Lintott, C., {et~al.} 2023, Monthly Notices of the Royal Astronomical Society, 521, 1775, \dodoi{10.1093/mnras/stad501}

\bibitem[{Heckman {et~al.}(1981)Heckman, T., M., Butcher, H., R., Miley, G., K., \& van}]{Heckman1981}
Heckman, T., M., {et~al.} 1981, ApJ, 247, 403

\bibitem[{Heckman {et~al.}(1984)Heckman, Miley, \& Green}]{Heckman1984}
Heckman, T.~M., Miley, G.~K., \& Green, R.~F. 1984, Astrophysical Journal, 281, 525

\bibitem[{Kim {et~al.}(2024)Kim, Gadotti, Querejeta, Pérez, Zurita, Neumann, van~de Ven, Méndez-Abreu, de~Lorenzo-Cáceres, Sánchez-Blázquez, Fragkoudi, Martins, Silva-Lima, Kim, \& Park}]{Kim2024}
Kim, T., Gadotti, D.~A., Querejeta, M., {et~al.} 2024, The Astrophysical Journal, 968, 87, \dodoi{10.3847/1538-4357/ad410e}

\bibitem[{Krajnovic {et~al.}(2006)Krajnovic, Cappellari, De~Zeeuw, \& Copin}]{Krajnovic2006}
Krajnovic, D., Cappellari, M., De~Zeeuw, P.~T., \& Copin, Y. 2006, Monthly Notices of the Royal Astronomical Society, 366, 787, \dodoi{10.1111/j.1365-2966.2005.09902.x}

\bibitem[{Lindblad(1963)}]{Lindblad1963}
Lindblad, B. 1963, Stockholms Observatoriums Annaler, 5

\bibitem[{Liu {et~al.}(2025)Liu, Li, \& Shen}]{Liu2025}
Liu, J., Li, Z., \& Shen, J. 2025, The Astrophysical Journal, 980, 146, \dodoi{10.3847/1538-4357/adabe0}

\bibitem[{Maller {et~al.}(2009)Maller, Berlind, Blanton, \& Hogg}]{Maller2009}
Maller, A.~H., Berlind, A.~A., Blanton, M.~R., \& Hogg, D.~W. 2009, The Astrophysical Journal, 691, 394, \dodoi{10.1088/0004-637X/691/1/394}

\bibitem[{Maschmann {et~al.}(2023)Maschmann, Halle, Melchior, Combes, \& Chilingarian}]{Maschmann2023}
Maschmann, D., Halle, A., Melchior, A.-L., Combes, F., \& Chilingarian, I.~V. 2023, A\&A, 670, A46, \dodoi{10.1051/0004-6361/202244746}

\bibitem[{Masters {et~al.}(2021)Masters, Krawczyk, Shamsi, Todd, Finnegan, Bershady, Bundy, Cherinka, Fraser-McKelvie, Krishnarao, Kruk, Lane, Law, Lintott, Merrifield, Simmons, Weijmans, \& Yan}]{Masters2021}
Masters, K.~L., Krawczyk, C., Shamsi, S., {et~al.} 2021, Monthly Notices of the Royal Astronomical Society, 507, 3923, \dodoi{10.1093/mnras/stab2282}

\bibitem[{Nevin {et~al.}(2016)Nevin, Comerford, Müller-S{\'{a}}nchez, Barrows, \& Cooper}]{Nevin2016}
Nevin, R., Comerford, J., Müller-S{\'{a}}nchez, F., Barrows, R., \& Cooper, M. 2016, The Astrophysical Journal, 832, 67, \dodoi{10.3847/0004-637x/832/1/67}

\bibitem[{Patel {et~al.}(2024)Patel, Arora, Courteau, Stone, Frosst, \& Widrow}]{Patel2024}
Patel, R., Arora, N., Courteau, S., {et~al.} 2024, The Astrophysical Journal, 972, 23, \dodoi{10.3847/1538-4357/ad58bc}

\bibitem[{Qiu {et~al.}(2024)Qiu, Shen, Feng, Chen, Chang, Zhao, \& Zeng}]{Qiu2024}
Qiu, J., Shen, S., Feng, S., {et~al.} 2024, The Astrophysical Journal, 976, 15, \dodoi{10.3847/1538-4357/ad6b8e}

\bibitem[{Rubinur {et~al.}(2016)Rubinur, Das, Kharb, \& Honey}]{Rubinur2016}
Rubinur, K., Das, M., Kharb, P., \& Honey, M. 2016, Monthly Notices of the Royal Astronomical Society, 465, 4772, \dodoi{10.1093/mnras/stw2981}

\bibitem[{S'anchez {et~al.}(2016)S'anchez, P'erez, S'anchez-Bl'azquez, Garc'ia-Benito, Ibarra-Mede, Gonz'alez, Rosales-Ortega, S'anchez-Menguiano, Ascasibar, Bitsakis, Law, Cano-d'iaz, L'opez-Cob'a, Marino, de~Paz, L'opez-S'anchez, Barrera-Ballesteros, Galbany, Mast, Abril-Melgarejo, \& Roman-Lopes}]{Sanchez2016}
S'anchez, S.~F., P'erez, E. A.~S., S'anchez-Bl'azquez, P., {et~al.} 2016, Revista Mexicana De Astronomia Y Astrofisica, 52, 171

\bibitem[{Schmidt {et~al.}(2019)Schmidt, Mast, Díaz, Agüero, Günthardt, Gimeno, Oio, \& Gaspar}]{Schmidt2019}
Schmidt, E.~O., Mast, D., Díaz, R.~J., {et~al.} 2019, The Astronomical Journal, 158, 60, \dodoi{10.3847/1538-3881/ab2882}

\bibitem[{Sormani {et~al.}(2024)Sormani, Sobacchi, \& Sanders}]{Sormani2024}
Sormani, M.~C., Sobacchi, E., \& Sanders, J.~L. 2024, Monthly Notices of the Royal Astronomical Society, 528, 5742, \dodoi{10.1093/mnras/stae082}

\bibitem[{Sparke \& Gallagher(2006)}]{Sparke2006}
Sparke, L.~S., \& Gallagher, J. S.~I. 2006, Galaxies in the Universe - 2nd Edition, by Linda S. Sparke and John S. Gallagher, III, pp. 406. Cambridge University Press, 2006

\bibitem[{Springel(2005)}]{Springel2005}
Springel, V. 2005, Monthly Notices of the Royal Astronomical Society, 364, 1105, \dodoi{10.1111/j.1365-2966.2005.09655.x}

\bibitem[{Stone {et~al.}(2022)Stone, Courteau, Arora, Frosst, \& Jarrett}]{Stone2022}
Stone, C., Courteau, S., Arora, N., Frosst, M., \& Jarrett, T. 2022, Astrophysical Journal Supplement Series, 262, \dodoi{10.3847/1538-4365/ac83ad}

\bibitem[{Tremaine \& Weinberg(1984)}]{Tremaine1984}
Tremaine, S., \& Weinberg, M. 1984, The Astrophysical Journal, 282, L5, \dodoi{10.1086/184292}

\bibitem[{Virtanen {et~al.}(2020)Virtanen, Gommers, Oliphant, Haberland, \& Mulbregt}]{Virtanen2020}
Virtanen, P., Gommers, R., Oliphant, T.~E., Haberland, M., \& Mulbregt, P.~V. 2020, Nature Methods, 17, 1

\bibitem[{Wang {et~al.}(2018)Wang, Luo, Song, Shen, Feng, Wang, Wang, Li, Du, Hou, Guo, Kong, \& Zhang}]{Wang2018}
Wang, M.-X., Luo, A.-L., Song, Y.-H., {et~al.} 2018, Monthly Notices of the Royal Astronomical Society, 482, 1889, \dodoi{10.1093/mnras/sty2818}

\bibitem[{Westfall {et~al.}(2019)Westfall, Cappellari, Bershady, Bundy, Belfiore, Ji, Law, Schaefer, Shetty, Tremonti, Yan, Andrews, Brownstein, Cherinka, Coccato, Drory, Maraston, Parikh, Sánchez-Gallego, Thomas, Weijmans, Barrera-Ballesteros, Du, Goddard, Li, Masters, Medel, Sánchez, Yang, Zheng, \& Zhou}]{Westfall2019}
Westfall, K.~B., Cappellari, M., Bershady, M.~A., {et~al.} 2019, The Astronomical Journal, 158, 231, \dodoi{10.3847/1538-3881/ab44a2}

\bibitem[{Willett {et~al.}(2013)Willett, Lintott, Bamford, Masters, Simmons, Casteels, Edmondson, Fortson, Kaviraj, Keel, Melvin, Nichol, Raddick, Schawinski, Simpson, Skibba, Smith, \& Thomas}]{Willett2013}
Willett, K.~W., Lintott, C.~J., Bamford, S.~P., {et~al.} 2013, Monthly Notices of the Royal Astronomical Society, 435, 2835, \dodoi{10.1093/mnras/stt1458}

\bibitem[{Xu \& Komossa(2009)}]{Xu2009}
Xu, D., \& Komossa, S. 2009, The Astrophysical Journal, 705, L20, \dodoi{10.1088/0004-637x/705/1/l20}

\bibitem[{Yazeedi {et~al.}(2021)Yazeedi, Katkov, Gelfand, Wylezalek, Zakamska, \& Liu}]{Yazeedi2021}
Yazeedi, A.~A., Katkov, I.~Y., Gelfand, J.~D., {et~al.} 2021, The Astrophysical Journal, 916, 102, \dodoi{10.3847/1538-4357/abf5e1}

\end{thebibliography}
\bibliographystyle{aasjournal}
\appendix

\section{Appendix}

\subsection{Examples of the rotation curve fittings\label{sec:flat}}
\setcounter{figure}{0}
\renewcommand{\thefigure}{A\arabic{figure}}

In this study, a tahn function (Equation \ref{gas rotation curve}) is used to fit the gas rotation curve of each barred galaxy using the $H\alpha$ velocities of the non-DPS spaxels. The best fits are obtained using the standard $\chi^2$ technique. 

\begin{figure*}
\centering
\includegraphics[width=.8\textwidth]{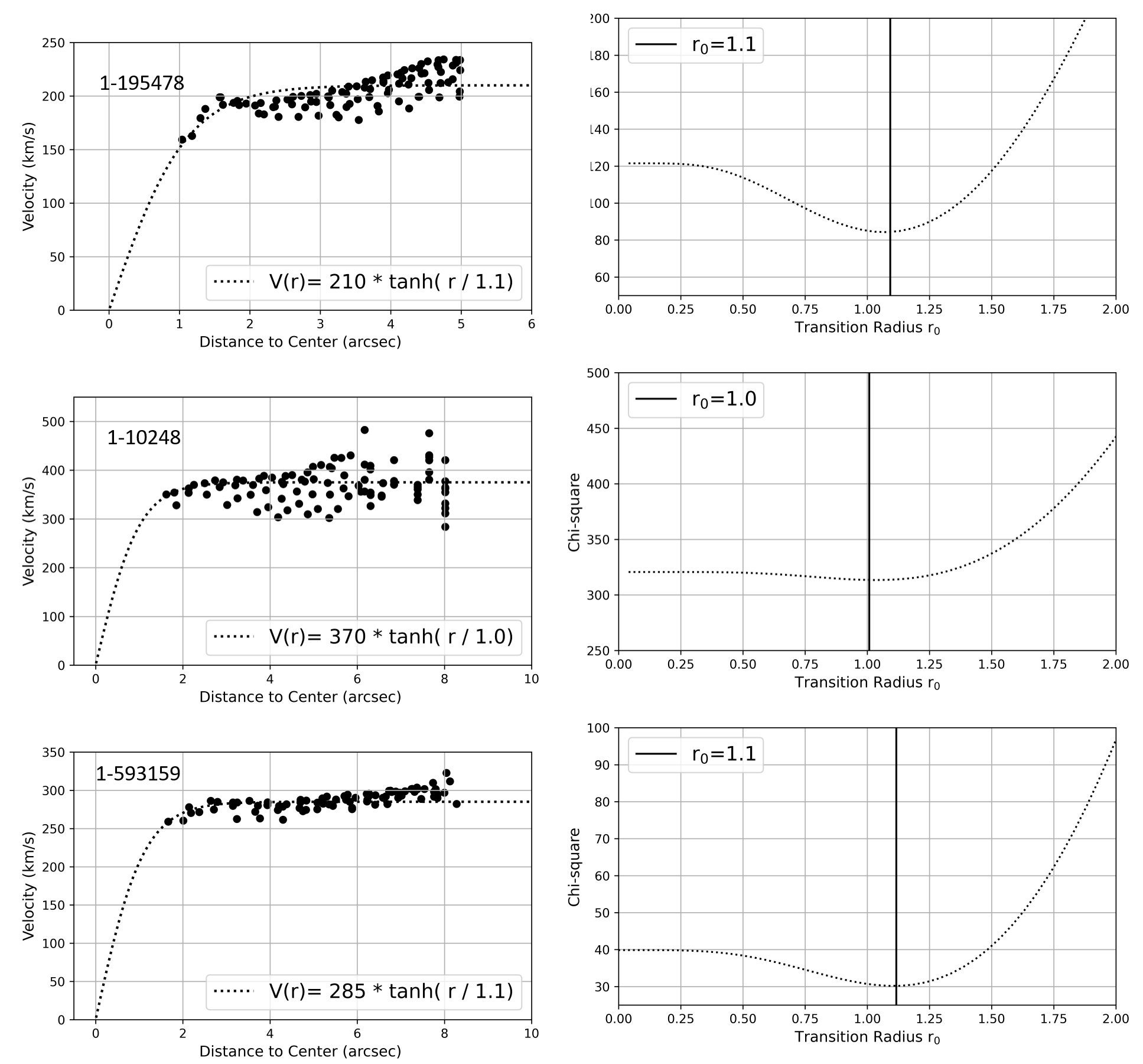}
\caption{\label{fig:f1} The rotation curve fitting process of three example galaxies: MaNGA-ID 1-195478 (top row), 1-10248 (middle row) and 1-593159 (bottom row). For each galaxy, the $H\alpha$ velocity data of non-DPS spaxels are shown in the left panels, while the $\chi^2$ values are plotted against the model parameter $r_0$ in the right panels. For the "flat sample" (MaNGA-ID 1-10248), its best fit of $r_0$ is obtaied from its upper limit.}
\end{figure*}

We show the details of the fitting for three example galaxies in Fig. \ref{fig:f1}, where the left panel shows the rotation curve data and the right panel shows the $\chi^2$ value as a function of the model parameter $r_0$. The top row shows a normal sample (MaNGA-ID 1-1-195478), whose $H\alpha$ velocity well covers the transition part of the rotation curve and the model parameter $r_0$ is well constrained. The middle row shows the case for a "flat sample" (MaNGA-ID 1-10248), where $H\alpha$ velocity data only covers the flat part of the rotation curve. For such a galaxy, only an upper limit of $r_0$ can be constrained. The bottom row shows that for our example DPG (MaNGA-ID 1-593159), whose transition radius $r_0$ is marginally constrained.

\subsection{Comparsion of rotation curve fitting using C97 model\label{sec:C97}}

In this study, we use the tanh function (Equation \ref{gas rotation curve}) to fit the gas rotation curves for simplicity. \citet{Stone2022} has suggested fitting the rotation curves of the central regions of barred galaxies with the C97 model \citep{Courteau1997},
\begin{equation}
V_{\rm{gas}}(r) = V_{\rm{gas,0}} \times \frac{(1+r_{\rm{gas,0}}/r)^{\beta}}{(1+(r_{\rm{gas,0}}/r)^{\gamma})^{1/\gamma}}, \label{c97}
\end{equation}
where $r_{\rm{gas,0}}$ is the turnover radius from the rising to flat regime, $V_{\rm{gas,0}}$ is the maximum velocity, $\beta$ controls the rising strength of the outskirts, and $\gamma$ determines the shape of the turnover to flat, respectively.

\begin{figure*}
\centering
\includegraphics[width=1\textwidth]{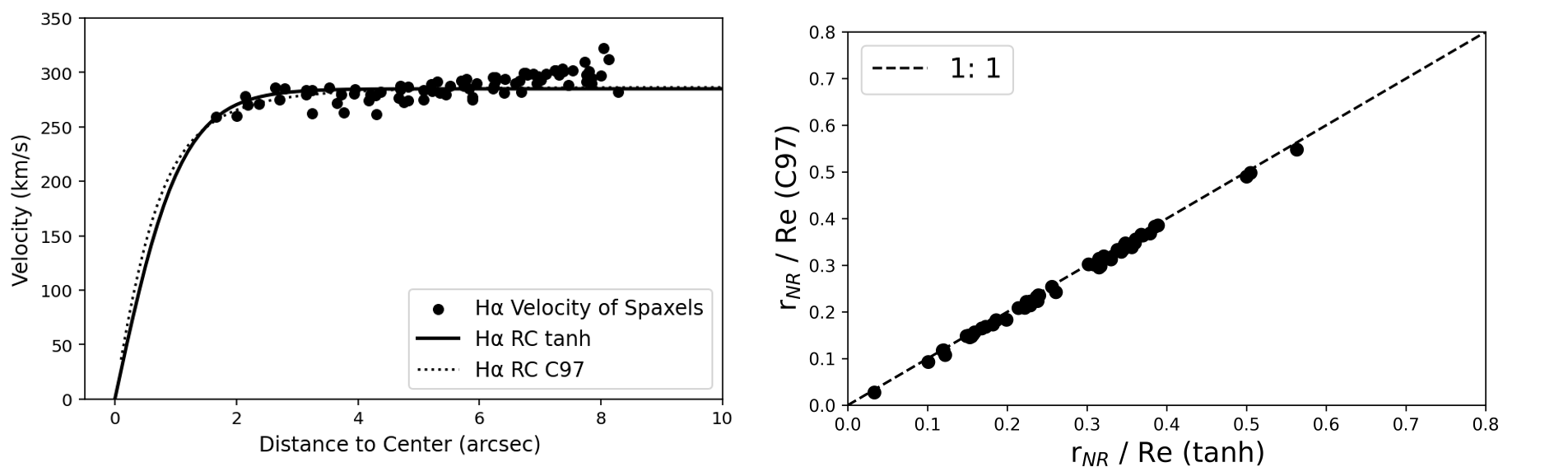}
\caption{\label{fig:f2} 
Left: The fitting of the gas rotatio curve $V_{\rm{gas}}(r)$ for the example DPG (MaNGA-ID 1-593159), where the solid line represents tanh model, while the dotted line is C97 model.
Right: The comaprsion of the $r_{\rm{NR}}$ values from tanh and C97 models. 
}
\end{figure*}

As a test, we make fittings of the rotation curves for all the barred DPGs. In the left panel of Fig. \ref{fig:f2}, we display the fitting results for the example DPG (MaNGA-ID 1-593159). The rotation curves derived from the two models exhibit minimal discrepancies.

Furthermore, we conduct a similar dynamical modeling of $r_{\rm{NR}}$ utilizing the fitted C97 rotation curves for all barred DPGs. The comparison between the resulting $r_{\rm{NR}}(C97)$ and those derived in section \ref{sec:Nuclear Rings of Barred DPGs} is illustrated in the right panel of Fig. \ref{fig:f2}. As shown by the plot, the two models are in very good agreement with each other.

\end{document}